\documentclass[12pt]{article}

\usepackage{amsmath, amssymb, slashed}
\usepackage[dvips]{graphicx}
\usepackage{epsf}
\usepackage{ulem}
\usepackage{url}
\usepackage{color}

\def\a{\alpha}

\def\e{\epsilon}

\def\g{\gamma}

\def\m{\mu}
\def\n{\nu}

\def\D{\Delta}
\def\G{\Gamma}

\newcommand{\MEV}{ {\rm MeV} }
\newcommand{\GEV}{ {\rm GeV} }
\newcommand{\TEV}{ {\rm TeV} }
\begin{document}

\begin{titlepage}
\begin{center}

\hfill ICRR-Report-641-2012-30 \\
\hfill IPMU-12-0231 \\
\hfill UT-12-44 \\
\hfill \today

\vspace{1.5cm}
{\large\bf Mass Splitting between Charged and Neutral Winos \\
at Two-Loop Level}
\vspace{2.0cm}

{\bf Masahiro Ibe}$^{(a, b)}$,
{\bf Shigeki Matsumoto}$^{(b)}$
and
{\bf Ryosuke Sato}$^{(b, c)}$

\vspace{1.0cm}
{\it
$^{(a)}${ICRR, University of Tokyo, Kashiwa, 277-8582, Japan} \\
$^{(b)}${Kavli IPMU, University of Tokyo, Kashiwa, 277-8583, Japan} \\
$^{(c)}${Department of Physics, University of Tokyo, Tokyo, 113-0033, Japan}
}
\vspace{2.0cm}

\abstract{
The recent result of the Higgs search 
at the LHC experiment
has lead to more attention to the supersymmetric standard models with heavy sfermions.
Among them, the models with the almost pure wino being the 
lightest supersymmetric particle (LSP) have been widely discussed 
due to their success in providing a consistent dark matter candidate.
The notable phenomenological feature of the wino LSP is the 
degeneracy with its charged SU(2)$_L$ partner (the charged wino) in mass.
The tiny mass splitting makes the charged wino long-lived,
which allows us to detect the wino production at the LHC experiment
by searching for the disappearing charged tracks inside the detectors.
Since the reach of the experiment is sensitive to the mass splitting,
it is mandatory to estimate it very precisely.  
We therefore perform a full calculation of the mass splitting at two-loop level, 
and find that the splitting is reduced by a few MeV compared to the one-loop calculation. 
This reduction leads to about a 10--30\,\% longer lifetime of the charged wino, 
with which the current constraint on the wino mass by the ATLAS experiment 
is improved by about 10\,\%.
}

\end{center}
\end{titlepage}
\setcounter{footnote}{0}
%%%%%%%%%%%%%%%%%%%%%%%%
%%%%% Introduction %%%%%
%%%%%%%%%%%%%%%%%%%%%%%%
\section{Introduction}
\label{sec: Introduction}
The supersymmetric standard model (SSM) is one of the most attractive candidates for physics beyond the standard model (SM). 
Both the discovery of the Higgs boson~\cite{:2012gk, :2012gu} and the null-observation of supersymmetry (SUSY) signals at the LHC experiment have given us some hints for SUSY model buildings.
In the minimal SSM (the MSSM), for example,
the Higgs boson mass is predicted to be smaller than the $Z$ boson mass at the tree-level. 
The observed mass of the Higgs boson 
at around 126\,GeV, 
therefore, indicates that huge radiative corrections to the Higgs self-coupling from the 
SUSY breaking effects are  required~\cite{Okada:1990vk}--\cite{Okada:1990gg}.

One of the simplest scenarios
leading to such huge corrections is putting the masses of 
 the sfermions (especially of the squarks) 
 at the scale of ${\cal O}(10$--$100)$ TeV~\cite{Okada:1990vk}--\cite{Giudice:2011cg}.  
It should be noted that 
although the squarks are far beyond the accessible range of the LHC experiment, in such cases,
this class of the scenarios 
 does not necessarily mean that all  the SUSY particles are as heavy as ${\cal O}(10$--$100)$ TeV. 
 For example, if we suppose that the SUSY breaking field is charged under some (gauge) symmetries,
gauginos cannot acquire their masses through the linear term of the SUSY breaking field in the gauge kinetic
functions of the MSSM. 
In this case,  the leading contributions to the gaugino masses come from the anomaly mediated 
contribution~\cite{Giudice:1998xp, Randall:1998uk}, which 
are one-loop suppressed compared to the squark masses. 
The gaugino masses are therefore predicted to be ${\cal O}(0.1$--$1)$\,TeV which 
are accessible at the LHC experiment. 
This class of the high-scale SUSY scenarios has recently attracted more attention, 
and phenomenological and cosmological aspects of the scenarios 
have been studied extensively~\cite{Ibe:2006de}--\cite{Asai:2008im}.%
\footnote{
In this Letter,
we base our discussion  on the pure gravity mediation scenarios~\cite{Ibe:2006de},
where the Higgsinos and the heavier Higgs bosons in the MSSM are predicted 
to be as heavy as the sfermions.
Our formulas obtained in the following analysis are  applicable to any heavy sfermion scenarios 
as long as the Higgsinos are much heavier than the gauginos as in
the scenarios such as PeV-Scale Supersymmetry\,\cite{Wells:2004di} 
and Spread Supersymmetry\,\cite{Hall:2011jd}.
} 

One of the most prominent features of the anomaly mediated gaugino spectrum is that 
the lightest supersymmetric particle (LSP) is predicted to be the almost pure neutral wino, 
which is highly degenerate with the charged wino in mass. 
This is because the mass splitting between these two particles is forbidden at the tree-level due to the 
approximate custodial symmetry. The mass splitting is dominated by the 
radiatively generated contributions 
which are estimated to be 160--170\,MeV at the 
one-loop level~\cite{Cheng:1998hc}--\cite{Gherghetta:1999sw}.

Due to the degeneracy between the charged and the neutral wino masses,
the charged wino decays mainly into the neutral wino and a soft pion with the decay length 
of $c\tau = {\cal O}(1$--$10)$\,cm, 
which allows the wino production to be detected by looking for a disappearing charged track inside
the detectors at the LHC experiment~\cite{ATLAS:2012jp}. 
This signal is characteristic for the high-scale SUSY scenarios with
the anomaly-mediated gaugino mass spectrum.
The expected number of the charged track is proportional to $\exp[-L/c\tau]$ with 
$L$ being the distance between a detector and a collision point, and hence, 
the reach of the experiment is very sensitive to $c\tau$. 
Therefore, a precise calculation of the decay length is mandatory.

In this article, we calculate the mass splitting between the charged and the neutral winos at the 
two-loop level. 
In Ref.\,\cite{Yamada:2009ve}, 
the splitting of the winos has been evaluated at the two-loop level 
in the heavy wino limit by calculating non-decoupling contributions. 
The result, however, cannot be directly applied to the wino mass
in the range of ${\cal O}(100)$\,GeV where the LHC experiment is searching for the winos.
We therefore perform a full two-loop calculation of the splitting 
including the non-decoupling effects. 
In the next section (section\,\ref{sec: mass-splitting}), 
we calculate the mass splitting at the two-loop level.
We will see that the contributions from the SM particles are dominant, 
while those from the SUSY particles are negligible. 
In section\,\ref{sec: lifetime}, we discuss the decay length of the charged wino 
and compared with the recent experimental results by the ATLAS collaboration\,\cite{ATLAS:2012jp}.
As a result, we find that the decay length of the charged wino
becomes 10--30\,\% longer than that obtained at the one-loop calculation.
This result makes the current constraint on the wino mass by the ATLAS experiment
severer than the LEP2 constraints\,\cite{Heister:2002mn}--\cite{LEP2}.
Section \ref{sec: summary} is devoted to summary of our discussion.

%%%%%%%%%%%%%%%%%%%%%%%%%%
%%%%% Mass splitting %%%%%
%%%%%%%%%%%%%%%%%%%%%%%%%%
\section{The mass splitting}
\label{sec: mass-splitting}
As mentioned above,
the neutral wino ($\tilde{\chi}^0$) and its charged SU(2)$_L$ partner (the charged wino, $\tilde{\chi}^\pm$) are almost degenerated in mass at the tree-level due to the approximate custodial symmetry. 
The dominant mass splitting,
$\delta m = m_{\tilde{\chi}^\pm} - m_{\tilde{\chi}^0}$,  is generated by radiative corrections, 
which  pick up the breaking of the custodial symmetry as pointed out in Ref. \cite{Gherghetta:1999sw}. 
In this section, we calculate the radiative corrections at the two-loop level. 

\subsection{SM contributions}
\label{subsec: SM-particles}
When the sfermions, Higgsinos, and the heavier Higgs bosons 
are in the range of ${\cal O}(10$--$100)$\,TeV and decouple
from the low energy physics below the TeV scale, 
 the neutral and the charged winos only couple to the 
 SM particles through the SU(2)$_L$ gauge interaction.
In such cases, the radiative correction to the mass splitting from the SM sector can be calculated 
by using
the  effective Lagrangian,
\begin{eqnarray}
{\cal L} &=&
{\cal L}_{\rm SM}
+
\frac{1}{2}
  \bar{\tilde{\chi}}^0 \left( i\slashed{\partial} - M_2 \right)\tilde{\chi}^0
+ \bar{\tilde{\chi}}^- \left( i\slashed{\partial} - M_2 \right)\tilde{\chi}^-
\nonumber \\
&&
- g \left( \bar{\tilde{\chi}}^0 \slashed{W}^\dagger \tilde{\chi}^- + h.c. \right)
+ g \bar{\tilde{\chi}}^-
\left( c_W \slashed{Z} + s_W \slashed{A} \right) \tilde{\chi}^-,
\end{eqnarray}
where ${\cal L}_{\rm SM}$ is the SM Lagrangian and $M_2$ is the invariant mass of the winos. 
The notation for the SM gauge fields is understood, 
and SU(2)$_L$ gauge coupling is denoted by $g$, while $c_W (s_W) = \cos \theta_W (\sin\theta_W)$ with $\theta_W$ being the weak mixing angle. 

The mass splitting  between the charged and the neutral winos is caused 
by the custodial symmetry breaking by U(1)$_Y$ gauge and Yukawa interactions.
It should be noted that the breaking of the custodial symmetry is highly suppressed 
at the tree-level in the wino-SM system.
In fact, at the tree-level, the breaking of the custodial symmetry is 
mediated through the Higgsino mixing.
As a result, the tree-level mass splitting is highly suppressed by the Higgsino mass, $\mu$, 
which is given by
\begin{eqnarray}
\delta m |_{\rm mixing} \simeq
\frac{ m_W^4 (\sin 2\beta)^2\tan^2 \theta_W }{ (M_1 - M_2) \mu^2 } \simeq
\frac{14~{\rm keV}}{\tan^2 \beta}
\left( \frac{300~{\rm GeV}}{M_1 - M_2}\right)
\left( \frac{100~{\rm TeV}}{\mu} \right)^2\ .
\label{eq: mixing}
\end{eqnarray}
Here, $m_W$ denotes the mass of the $W$-boson, $\beta$ the Higgs mixing angle of the MSSM, 
and $M_1$ the mass of the bino.% 
\footnote{The mass splitting in Eq.\,(\ref{eq: mixing}) is valid for $M_1-M_2\gg m_Z$.}
As we will see below, the above tree-level mass splitting is sub-dominant 
compared to the radiatively generated mass splitting.\footnote{
In the Split Supersymmetry models \cite{Split} where the Higgsino can be as light as the gauginos, the tree-level contribution to the mass splitting is not necessarily negligible. 
}

\subsubsection{The pole mass}
The pole mass of a spin half particle can be extracted from the $1$PI effective
two-point function,
\begin{eqnarray}
\G_2 =  \slashed{p} - M_0+ \Sigma_K(p^2) \slashed{p} + \Sigma_M(p^2)\ ,
\end{eqnarray}
 with $p$ being the four momentum of the particle and $M_0$ the tree-level mass.
Thus, for given self-energy functions, $\Sigma_K$ and $\Sigma_M$,
the pole mass is iteratively given by
\begin{eqnarray}
M_{\rm pole} =
{\rm Re}
\left[
\frac{ M_0 - \Sigma_M(M_{\rm pole}^2) }{1 + \Sigma_K(M_{\rm pole}^2)}
\right] \ .
\label{eq: def_of_pole_mass}
\end{eqnarray}
In a perturbative analysis, we expand the above pole mass as a power series of coupling constants.
At the two-loop level, the above iterative expression of the pole mass  
is reduced to 
\begin{eqnarray}
M_{\rm pole} &=&
{\rm Re}
\Biggl[ M_0 - \Sigma_M^{(1)} - M_0 \Sigma_K^{(1)}
            - \Sigma_M^{(2)} - M_0 \Sigma_K^{(2)}
\nonumber \\
&& \qquad
+ \left( \Sigma_M^{(1)} + M_0 \Sigma_K^{(1)} \right)
  \left(
  \Sigma_K^{(1)} + 2M_0\dot{\Sigma}_M^{(1)} + 2M_0^2 \dot{\Sigma}_K^{(1)} 
  \right) \Biggr]_{p^2 = M_0^2}\ .
\label{eq: pole_mass}
\end{eqnarray}
Here,  $\Sigma_{K, M}^{(1)}$ and $\Sigma_{K, M}^{(2)}$ are 
the self-energy functions at the one- and two-loop levels, respectively, 
while the dotted functions, 
$\dot{\Sigma}_{K, M}^{(1)}$, denote the derivatives of $\Sigma_{K, M}^{(1)}$ with respect to $p^2$.

\subsubsection{Renormalization scheme and input parameters}
We take the input parameters to the above effective Lagrangian:
\begin{eqnarray}
\hat{\alpha},
~\hat{m}_W,
~\hat{m}_Z,
~\hat{M}_2,
~\hat{m}_t,
~\hat{m}_h,
~{\rm and}
~Q,
\nonumber
\end{eqnarray}
where the hatted variables denote the  $\overline{\rm MS}$ variables, and $Q$ is the renormalization scale. 
All the quark and lepton masses except for the top quark mass are neglected in our analysis.

To relate the above listed input parameters (the $\overline{\rm MS}$ variables) to
the experimental observables,
we have to take finite renormalization effects into account. 
In the following analysis, we extract the input parameters by using the 
 renormalized relations at the one-loop level,
\begin{eqnarray}
\hat{\alpha}^{-1}_{\rm SM} &=& \hat{\alpha}^{-1}
\left[1 + \tilde{\Pi}^{(\tilde{\chi})}_{\gamma\gamma}(Q^2)/Q^2 \right]\ ,
\label{eq: alpha-MS-bar} \\
m_W^2 &=& \hat{m}_W^2 - \Pi_{WW}(m_W^2)\ , \\
m_Z^2 &=& \hat{m}_Z^2 - \Pi_{ZZ}(m_Z^2)\ , 
\label{eq: mZ-MS-bar} 
\\
m_{\tilde{\chi}^0} &=& \hat{M}_2
- \hat{M}_2 \Sigma_K^{(1)}(m_{\tilde{\chi}^0}^2)
- \Sigma_M^{(1)}(m_{\tilde{\chi}^0}^2)\ ,
\label{eq: mchi0}
\end{eqnarray}
where all the self-energies ($\Pi_{xx}$ and $\Sigma^{(1)}_{K,M}$) 
used in the above equations are given in the appendix~\ref{sec: one-loop-corrections}. 
Here, $\hat\alpha_{\rm SM}^{-1}$ denotes the QED fine structure constant in the $\overline{\rm MS}$
scheme in the SM at the $Z$-boson mass scale,
$m_{W,Z}$ the physical $W$ and $Z$ boson masses,
$m_{\tilde\chi^0}$  the physical neutral wino mass.
It should be noted that the one-loop relations are precise enough for the two-loop estimation
of the wino mass splitting, since the leading mass splitting starts at the one-loop level.

The top quark and the Higgs boson appear only at the two-loop calculation of the mass splitting. 
Thus, the $\overline{\rm MS}$ variables $\hat{m}_t$ and $\hat{m}_h$ may be replaced 
with their physical masses $m_t$ and $m_h$ at this level of precision.
As for the top quark mass, however, we use the $\overline{\rm MS}$ top mass 
at the one-loop level for $\hat{m}_t$.%
\footnote{The finite renormalization effect connecting between $\hat{m}_t$ ($\overline{\rm MS}$ mass) 
and $m_t$ (pole mass) is the same as those in the SM, 
because the scalar top quarks are heavy and decoupled.} 
As we will see, the $Q$ dependence of the mass splitting at the two-loop level comes 
mainly from those of the top mass $\hat{m}_t$.
We set, on the other hand, $\hat{m}_h = m_h$ since the running of the Higgs mass 
does not cause significant effects on the splitting. 

Once we obtain the input parameters, $\hat{\alpha}$, $\hat{m}_W$, and $\hat{m}_Z$
from Eqs.\,(\ref{eq: alpha-MS-bar})-(\ref{eq: mZ-MS-bar}), 
we can calculate $\hat{g}$, $\hat{g}'$ using tree-level relations. 
In deriving the one-loop relations in  Eqs.\,(\ref{eq: alpha-MS-bar})-(\ref{eq: mchi0}),
we also obtain the counter-terms to subtract ultra-violet (UV) divergences. 
These counter-terms play important roles to calculate $\Sigma_{K,M}^{(2)}$, as will be discussed later.

\subsubsection{The mass splitting at one-loop level}
%%%%%%%%%%%%%%%%%%%%%%%%%%%%%%%%%%%%%%%%%%%%%%%%%%
\begin{figure}[t]
\begin{center}
\includegraphics[width=0.31\hsize]{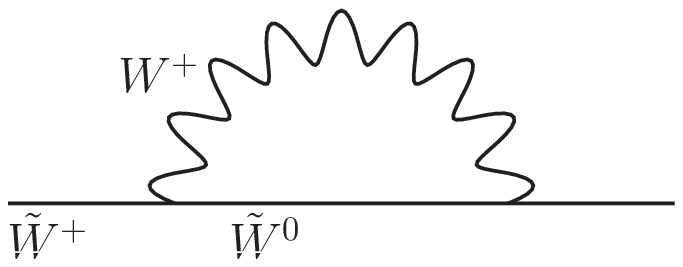}
~~
\includegraphics[width=0.31\hsize]{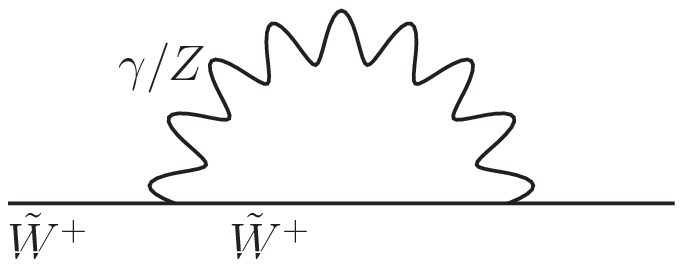}
~~
\includegraphics[width=0.31\hsize]{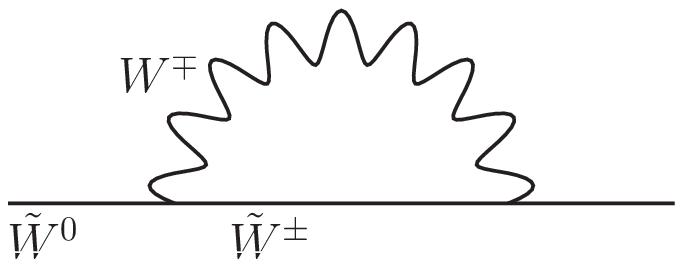}
\caption{\small \sl One-loop diagrams contributing to the functions $\Sigma_{M,K}^{(1)}$ in Eq.\,(\ref{eq: pole_mass}).}
\label{fig: 1-loop}
\end{center}
\end{figure}
%%%%%%%%%%%%%%%%%%%%%%%%%%%%%%%%%%%%%%%%%%%%%%%%%%
The one-loop result of the mass splitting between neural and charged winos is well known~\cite{Cheng:1998hc}--\cite{Gherghetta:1999sw} and used in the earlier literature.
The loop diagrams of the winos and gauge bosons shown in Fig.\,\ref{fig: 1-loop} lead 
to the functions $\Sigma_{K,M}^{(1)}$. 
With the use of the formula in Eq.\,(\ref{eq: pole_mass}) and the self-energies 
$\Sigma_{K,M}^{(1)}$ given in the appendix\,\ref{sec: one-loop-corrections}, 
the mass splitting $\delta m = m_{\tilde{\chi}^\pm} - m_{\tilde{\chi}^0}$ 
at the one-loop level is given by
\begin{eqnarray}
\delta m &=&
-\hat{M}_2 \Sigma_{K, \pm}^{(1)}(\hat{M}_2^2)
-\Sigma_{M,\pm}^{(1)}(\hat{M}_2^2)
+\hat{M}_2 \Sigma_{K,0}^{(1)}(\hat{M}_2^2)
+\Sigma_{M,0}^{(1)}(\hat{M}_2^2)
\nonumber \\
&=&
(\hat{g}^2 \hat{M}_2/8\pi^2)
[f(\hat{m}_W^2/\hat{M}_2^2)-\hat{c}_W^2 f(\hat{m}_Z^2/\hat{M}_2^2)], \label{eq:oneloop}
\rule{0mm}{5mm}
\end{eqnarray}
where the function $f(z)$ is defined as $f(z) = \int_0^1 dx (1+x) \log[1 + z(1-x)/x^2]$. 
In the heavy wino limit, $\hat{M}_2 \gg$ $\hat{m}_{Z, W}$,
the mass splitting is reduced to
\begin{eqnarray}
\delta m \simeq \frac{\hat{g}^2}{8\pi}(\hat{m}_W - \hat{c}_W^2 \hat{m}_Z)\ ,
\end{eqnarray}
which is about $160$--$170$\,MeV.

%%%%%%%%%%%%%%%%%%%%%%%%%%%%%%%%%%%%%%%%%%%%%%%%%%
\begin{figure}[htpb]
\begin{minipage}{\hsize}
\begin{center}
  \begin{minipage}{0.32\hsize}
  \begin{center}
    \includegraphics[width=4cm]{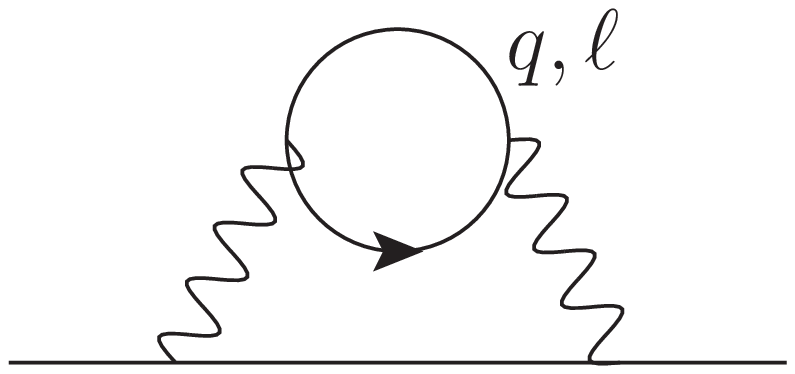}\\ (a)
  \end{center}
  \end{minipage}
  \begin{minipage}{0.32\hsize}
  \begin{center}
    \includegraphics[width=4cm]{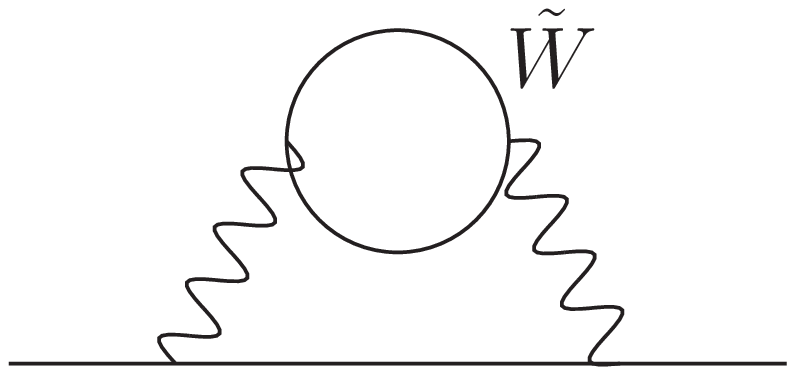}\\ (b)
  \end{center}
  \end{minipage}
  \begin{minipage}{0.32\hsize}
  \begin{center}
    \includegraphics[width=4cm]{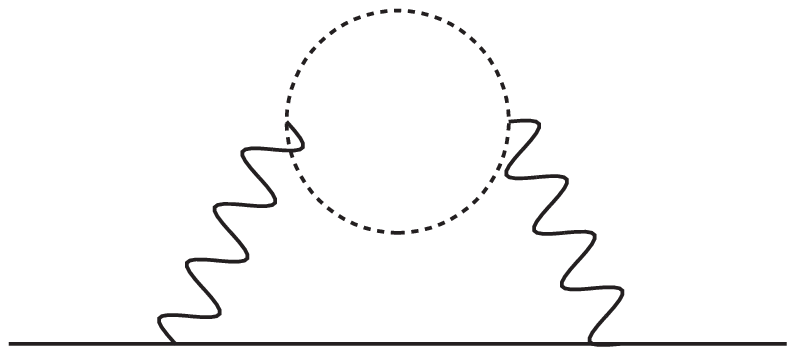}\\ (c)
  \end{center}
  \end{minipage}
  \\%
  \begin{minipage}{0.32\hsize}
  \begin{center}
    \includegraphics[width=4cm]{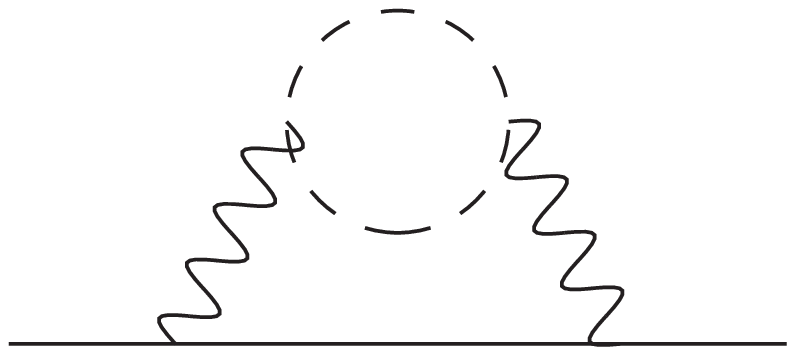}\\ (d)
  \end{center}
  \end{minipage}
  \begin{minipage}{0.32\hsize}
  \begin{center}
    \includegraphics[width=4cm]{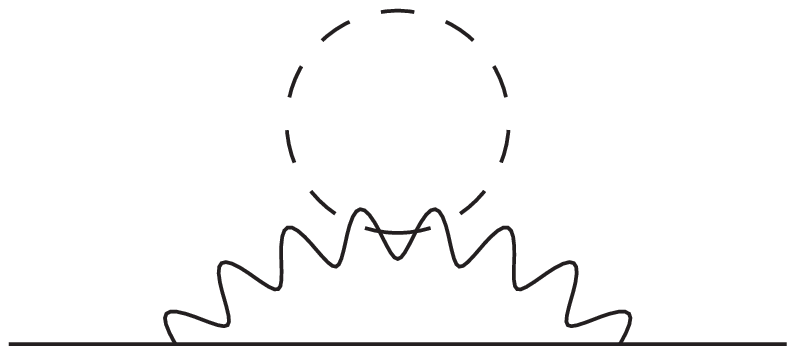}\\ (e)
  \end{center}
  \end{minipage}
  \begin{minipage}{0.32\hsize}
  \begin{center}
    \includegraphics[width=4cm]{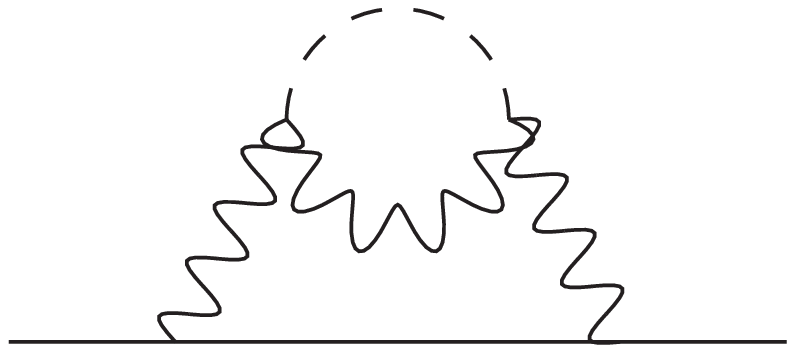}\\ (f)
  \end{center}
  \end{minipage}
  \\%
  \begin{minipage}{0.32\hsize}
  \begin{center}
    \includegraphics[width=4cm]{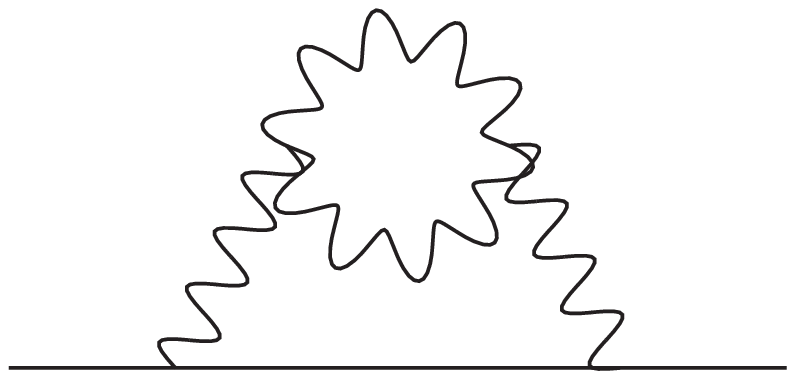}\\ (g)
  \end{center}
  \end{minipage}
  \begin{minipage}{0.32\hsize}
  \begin{center}
    \includegraphics[width=4cm]{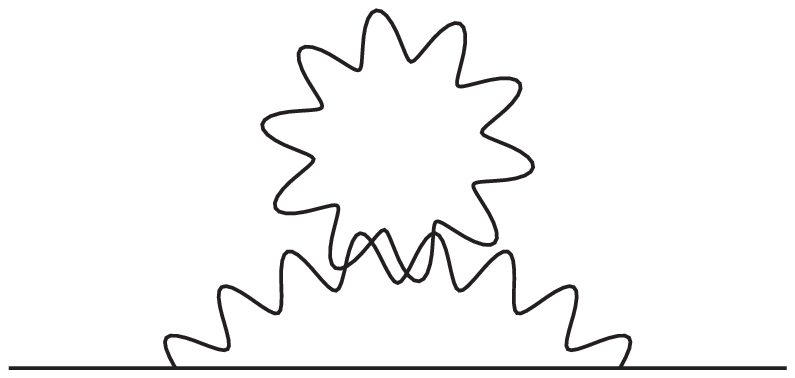}\\ (h)
  \end{center}
  \end{minipage}
  \begin{minipage}{0.32\hsize}
  \begin{center}
    \includegraphics[width=4cm]{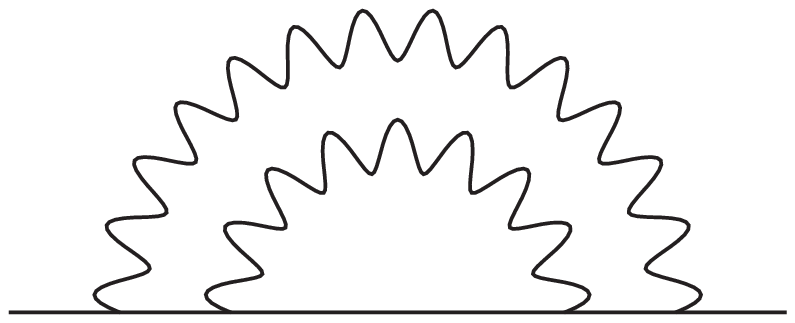}\\ (i)
  \end{center}
  \end{minipage}
  \\%
  \begin{minipage}{0.32\hsize}
  \begin{center}
    \includegraphics[width=4cm]{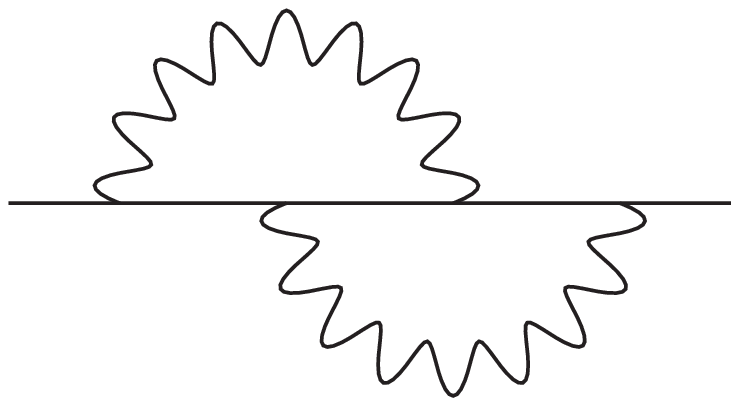}\\ (j)
  \end{center}
  \end{minipage}
  \begin{minipage}{0.32\hsize}
  \begin{center}
    \includegraphics[width=4cm]{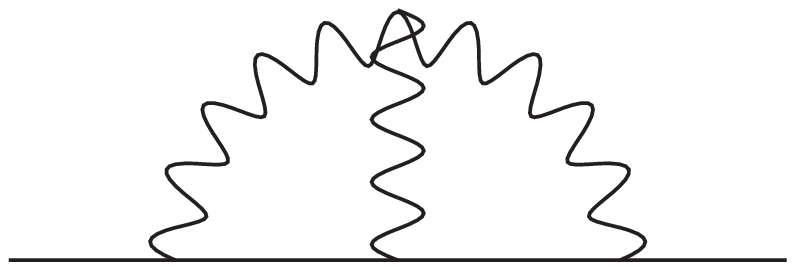}\\ (k)
  \end{center}
  \end{minipage}
  \caption{\small \sl Two-loop diagrams contributing to the functions $\Sigma_{M,K}^{(2)}$ in 
  Eq.\,(\ref{eq: pole_mass}). Diagram (a) includes the SM fermion loops, while (b) includes the wino loop. Diagram (c) includes the Faddeev-Popov ghost loop, and (d--f) includes the SM Higgs loop.}
\label{fig: 2-loop}
\end{center}
\end{minipage}
\vspace{2cm}\\
\begin{minipage}{\hsize}
\begin{center}
  \begin{minipage}{0.45\hsize}
  \begin{center}
    \includegraphics[width=5cm]{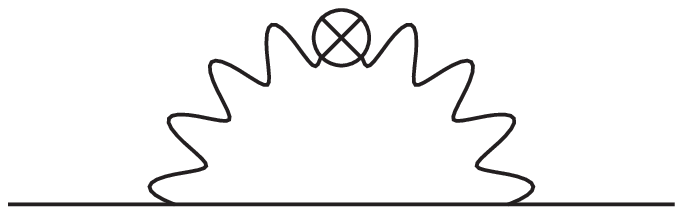}\\ (a)
  \end{center}
  \end{minipage}
  \begin{minipage}{0.45\hsize}
  \begin{center}
    \includegraphics[width=5cm]{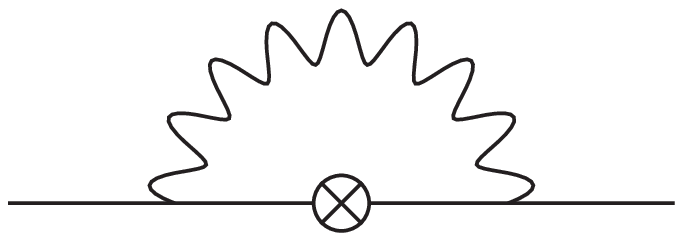}\\ (b)
  \end{center}
  \end{minipage}
  \\%
  \begin{minipage}{0.45\hsize}
  \begin{center}
    \includegraphics[width=5cm]{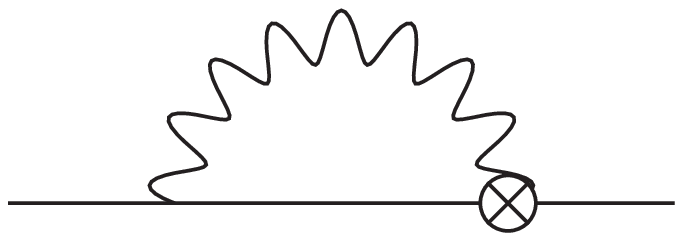}\\ (c)
  \end{center}
  \end{minipage}
  \begin{minipage}{0.45\hsize}
  \begin{center}
    \includegraphics[width=5cm]{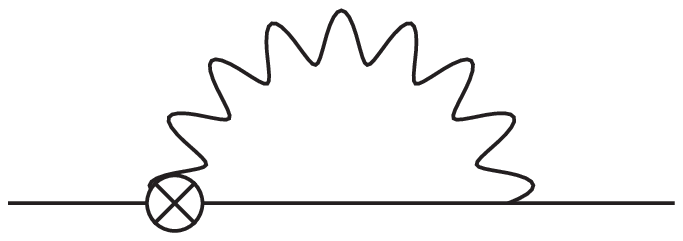}\\ (d)
  \end{center}
  \end{minipage}
  \caption{\small \sl Diagrams including counter-terms which contribute to the function $\Sigma_{M,K}^{(2)}$ in Eq.\,(\ref{eq: pole_mass}). The counter-terms are determined to renormalize one-loop divergences.}
\label{fig: 1-loop-ct}
\end{center}
\end{minipage}
\end{figure}
%%%%%%%%%%%%%%%%%%%%%%%%%%%%%%%%%%%%%%%%%%%%%%%%%%
\subsubsection{The mass splitting at two-loop level (strategy)}
The two-loop self-energies, $\Sigma_{K,M}^{(2)}(\hat{M}_2^2)$,
are obtained from the two-loop diagrams  (Fig.\,\ref{fig: 2-loop}) and from the diagrams including counter-terms which cancel the one-loop UV divergences  (Fig.\,\ref{fig: 1-loop-ct}). 
In our actual analysis,  we  first calculated the two-loop 1PI amplitudes using 
{\tt FeynArts}~\cite{Hahn:2000kx} and {\tt FeynCalc}~\cite{Mertig:1990an},
which were reduced to a set of basis integrals by {\tt TARCER}~\cite{Mertig:1998vk}. 
We finally evaluated the integrals numerically using {\tt TSIL}~\cite{Martin:2005qm}. 
For the diagrams including counter-terms, we used the ones given in Appendix~\ref{sec: counter}.
As a nontrivial cross check, we have confirmed that all the UV divergences are properly canceled.

We also have to care about infra-red (IR) singularities. 
For the charged wino, the amplitude in Fig.\,\ref{fig: 2-loop}-(i) in which a photon is circulating
in the outer loop and the one in Fig.\,\ref{fig: 1-loop-ct}-(b) with the photon loop behave as
\begin{eqnarray}
\Sigma_{K,M}^{(2)}(p^2=\hat{M}_2^2) \sim
\int \frac{d^4 k}{(2\pi)^4} \frac{1}{(k\cdot p)^2} \frac{1}{k^2} \ ,
\end{eqnarray}
and hence, they are IR divergent.
In addition, the derivatives, $\dot{\Sigma}_{K,M}^{(1)} |_{p^2=\hat{M}_2^2}$, 
are also IR divergent due to the diagram including a photon propagator. 
We have checked that all the IR divergences are canceled with each other 
when we evaluate the pole mass in Eq.\,(\ref{eq: pole_mass}).
See the appendix~\ref{app: IR cancel} for more discussions on the cancellation of the IR divergences. 

%%%%%%%%%%%%%%%%%%%%%%%%%%%%%%%%
\subsubsection{The mass splitting at two-loop level (result)}
\begin{figure}[p]
\begin{center}
  \includegraphics[width=0.8\hsize]{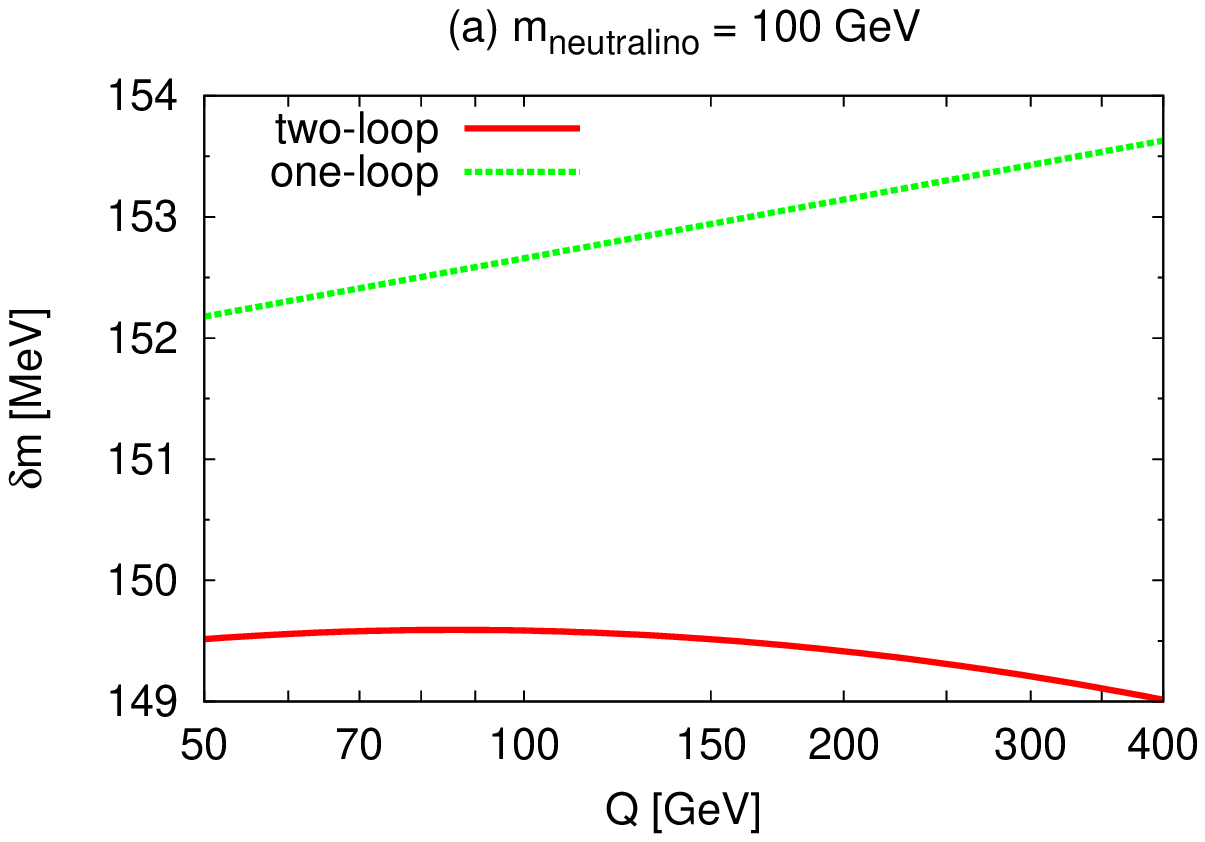} \vspace{1cm}\\
  \includegraphics[width=0.8\hsize]{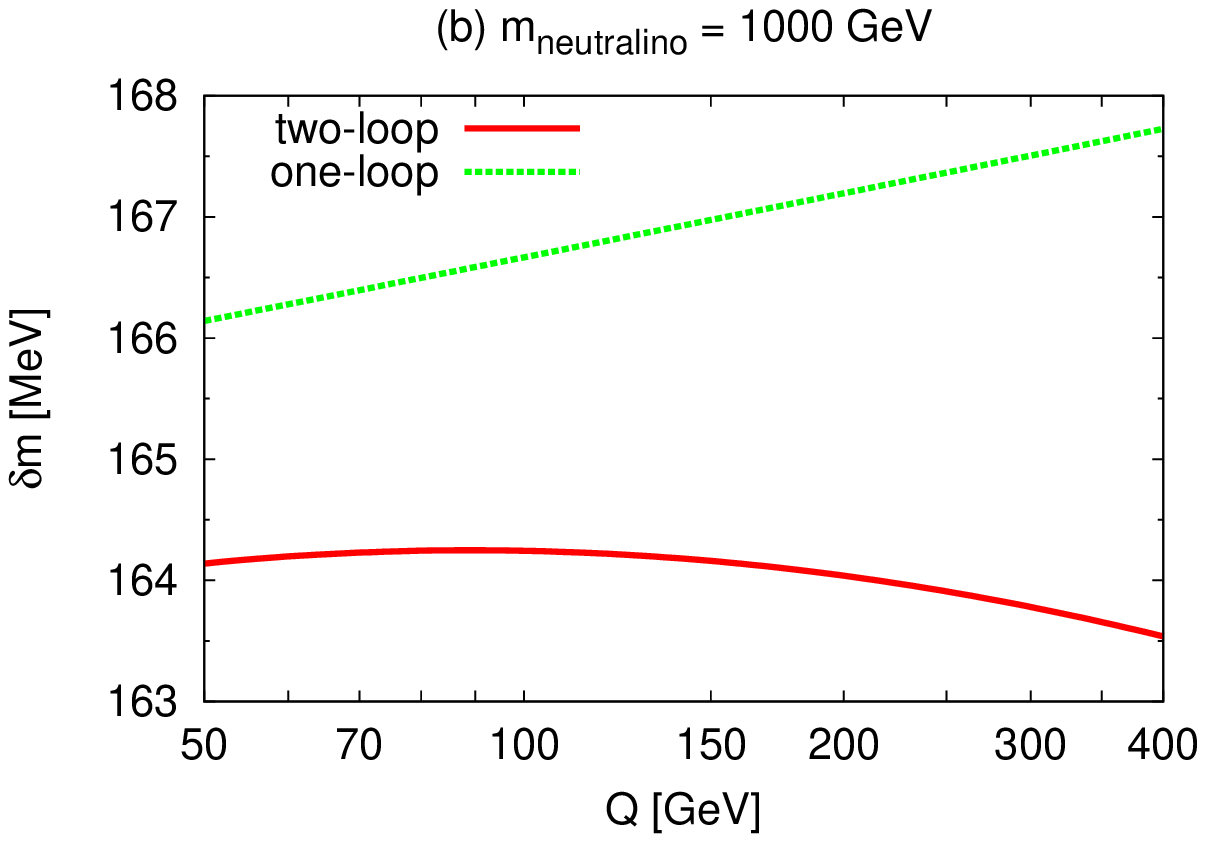}
  \caption{
{\sl
The renormalization scale dependence of $\delta m$.
The green lines show $\delta m$ at one-loop level in Eq.\,(\ref{eq:oneloop}),
and the red lines is $\delta m$ at the two-loop level which is evaluated by Eq.\,(\ref{eq: pole_mass}) 
in $\overline{\rm MS}$ scheme.
We take $m_{\tilde\chi^0} = $100\,GeV (a) and 1000\,GeV (b).
Here, we take $\hat m_t(m_t) = $163.3\,GeV and $m_h =$125.5\,GeV.
  }}\label{fig:qdep}
\end{center}
\end{figure}
%%%%%%%%%%%%%%%%%%%%%%%%%%%%%%%%

Now, let us show the resultant mass splitting at the two-loop level.
In the following, 
we take $\hat \alpha_{\rm SM}^{-1}(m_Z) = 127.944 \pm 0.014$, $m_W = 80.385\pm0.015\,$GeV, $m_Z = 91.1876\pm0.0021\,$GeV \cite{Beringer:1900zz},
$\hat m_t(m_t) = 163.3\pm2.7\,$GeV \cite{Alekhin:2012py} and
$m_h = 125.5\pm 0.7\,$GeV
as the SM input parameters.%
\footnote{
The ATLAS collaboration reports the Higgs boson mass as 
$125.2 \pm 0.3 \pm 0.6$\,GeV~\cite{atlas-conf-2012-170},
and CMS collaboration reports $125.8 \pm 0.4 \pm 0.4\,$GeV~\cite{cms-pas-hig-12-045}.
Here, we take naive average over these results, 
and combine statistical and systematic error.}

In Fig.\,\ref{fig:qdep}, 
we show the renormalization scale  dependence of $\delta m$,
which is the dominant source of the theoretical uncertainty of the mass splitting
at the two-loop level.
In the figure, the two-loop result is numerically evaluated by Eq.\,(\ref{eq: pole_mass})
in the $\overline{\rm MS}$ scheme, 
while the one-loop result is evaluated by Eq.\,(\ref{eq:oneloop}) 
in terms of the on-shell $W$ and $Z$ boson masses, i.e., 
\begin{eqnarray}
\label{eq: 1-loop_OS}
\delta m_{\rm 1loop} = \frac{\hat\a_{\rm SM}(Q)}{2\pi \tilde s_W^2}\left[ f(m_W^2/M_2^2) - \tilde c_W^2 f( m_Z^2/M_2^2)\right],
\end{eqnarray}
where we defined $\tilde c_W^2 = m_W^2 / m_Z^2$ and $\tilde s_W^2 + \tilde c_W^2 = 1$.
The $Q$ dependence of the one-loop result in Eq.\,(\ref{eq: 1-loop_OS})
comes from the running of the gauge coupling constant,
while
the $Q$ dependence of the two-loop result in Eq.\,(\ref{eq: pole_mass}) comes from all the $\overline{\rm MS}$ parameters.
The figure shows that the $Q$ dependence becomes weaker 
at the two-loop level as  expected, since the mass splitting should not depend on $Q$ 
at full order.
In our analysis, we found that the dominant source of the $Q$ dependence of the two-loop result
is the running of the top quark mass.

The uncertainty of the mass splitting due to the choice of $Q$ is expected 
to be compensated by the three-loop contributions including the QCD and the top-Yukawa interactions.
These corrections are generated by the diagrams including top-quark loop,
then, it is expected to be small if we take the renormalization scale as the top-quark mass.
For this reason, we fix the renormalization scale as $Q = \hat m_t$ in our calculation.
The $Q$ dependence of the two-loop result 
gives us a rough estimation of the uncertainty of the mass splitting
from the higher-loop effects.
We estimate the uncertainty of the mass splitting due to the choice of $Q$ by
%%%%%%%%%%%%%%%%%%%%%%%%%%%%%%%%%%%%%%%%%%%%%%%%%%%%%%%%%%%%%%%%%%%%%%%%%%%%%%%%
\begin{eqnarray}
\label{eq:Qdep}
{\mit \D}_Q \delta m &=& \frac{d\delta m}{d\log Q} \biggr|_{Q=\hat m_t}\ . \label{eq:uncertaintyQ}
\end{eqnarray}

In addition to the above uncertainty,
there are expected to be other uncertainties from the higher-loop corrections 
which are not encapsulated in the choice of the renormalization scale. 
At the three-loop level, for example, the dominant non-decoupling contribution to the 
mass splitting is expected to be proportional to $m_t$ and the QCD coupling.
Although the numerical factors of those corrections cannot be determined 
unless explicitly calculated, we give naive estimations to those higher loop
corrections by 
\begin{eqnarray}
{\mit\D}_{\rm 3-loop} \delta m
= \left(\frac{\a_2}{4\pi}\right)^2\left(\frac{\a_s}{4\pi}\right) \pi m_t \simeq 0.033\,\MEV, \label{eq:threshold_twoloop}
\end{eqnarray}
where $\a_2 = g^2 / 4\pi$.
Here, we have multiplied a factor of $\pi$ which is expected to accompany the 
non-decoupling effects at $M_2 \gg m_t$.
\footnote{
We have confirmed that the naive estimation of the two-loop contribution,
\begin{eqnarray}
{\mit\D}_{\rm 2-loop}\delta m = \left(\frac{\a_2}{4\pi}\right)^2 \pi m_t \simeq 3.9~\MEV\ , \label{eq:threshold_oneloop} 
\end{eqnarray}
gives a fair estimation of our two-loop numerical results.
}

The experimental errors of the input parameters also 
lead to uncertainties of the mass splitting.
As we summarize in Tab.\,\ref{tab:error}, however, 
the  effects of the experimental errors are relatively small compared to
the theoretical errors.
As a result, we find that the uncertainty on $\delta m$ is dominated 
by the three loop logarithmic corrections, i.e. the renormalization scale dependence.
\begin{table}
\begin{center}
\begin{tabular}{ccc}
Type of error & Estimate of the error & Impact on $\delta m$\\
\hline
$\hat\a_{\rm SM}(m_Z)$  & experimental uncertainty in $\hat\a_{\rm SM}(m_Z)$  & $\pm 0.018~\MEV$\\
$m_W$ & experimental uncertainty in $m_W$ & $\pm 0.019~\MEV$\\\
$m_Z$ & experimental uncertainty in $m_Z$ & $\pm 0.001~\MEV$\\
$\hat m_t$ & experimental uncertainty in $\hat m_t$ & $\pm 0.081~\MEV$\\
$m_h$ & experimental uncertainty in $m_h$ & $\pm 0.002~\MEV$\\
{\bf Experiment} & {\bf Total combined in quadrature} & $\pm 0.085~\MEV$\\
\hline
choice for $Q$ & QCD and top Yukawa at one-loop by Eq. (\ref{eq:uncertaintyQ}) & $\pm (0.3-0.4)~\MEV$\\
three-loop & naive estimation by Eq. (\ref{eq:threshold_twoloop}) & $\pm 0.033~\MEV$\\
{\bf Theory} & {\bf Total combined in quadrature} & $\pm (0.3-0.4)~\MEV$\\
\hline
\hline
{\bf Total} & {\bf Total combined in quadrature} & $\pm (0.31-0.41)~\MEV$
\end{tabular}
\caption{Experimental and theoretical errors in the evaluation on $\delta m$ at two-loop level.}\label{tab:error}
\end{center}
\end{table}

\begin{figure}[p]
\begin{center}
\includegraphics[width=\hsize]{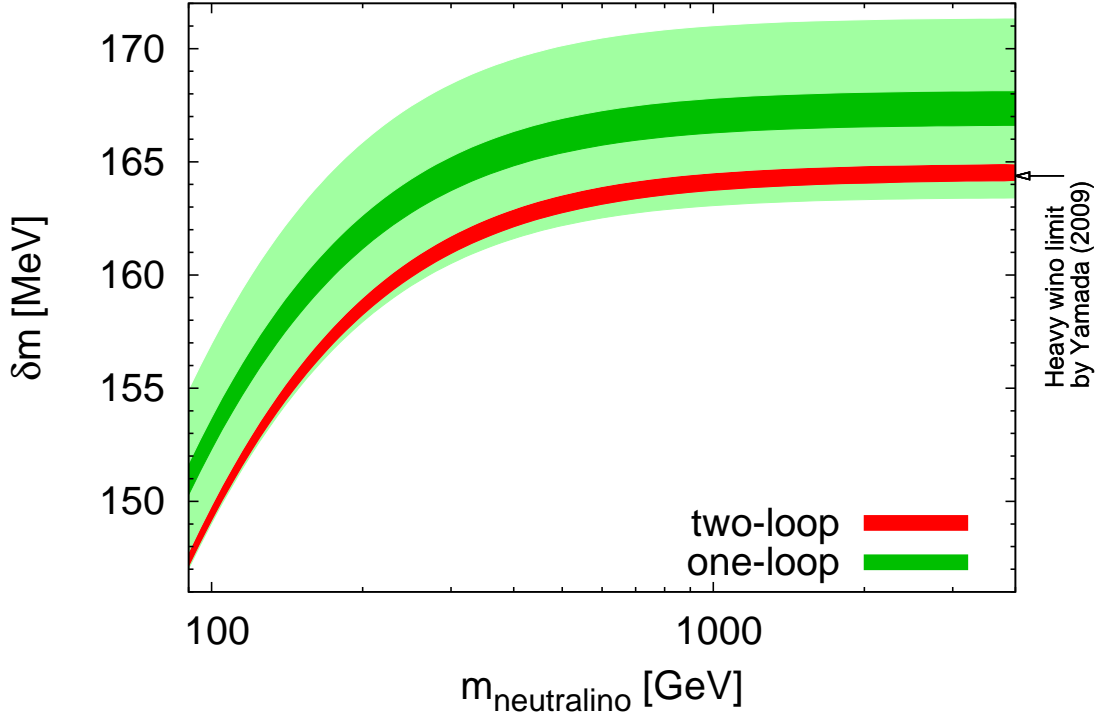}
\caption{\sl{
The wino mass splitting $\delta m$ as a function of $m_{\tilde\chi^0}$.
The dark green band shows $\delta m$ at the one-loop level which is evaluated by Eq.\,(\ref{eq:oneloop}) 
with uncertainty induced by $Q$ dependence, and the red band shows 
$\delta m$ at two-loop which is evaluated by Eq.\,(\ref{eq: pole_mass}) in $\overline{\rm MS}$ scheme.
The light green band shows the uncertainty for one-loop result evaluated by Eq.\,(\ref{eq:threshold_oneloop}).
The uncertainties for the two-loop result induced by the SM input parameters 
and the non-logarithmic corrections are negligible (see Tab.\,\ref{tab:error}).
An arrow shows the result of Ref.\,\cite{Yamada:2009ve},
which is given by $\delta m = 164.4~\MEV$ for $m_h = 125~{\rm GeV}$ and $m_t = 163.3~{\rm GeV}$.
}
}\label{fig:dm}
\end{center}
\end{figure}

In Fig.\,\ref{fig:dm},
we show the mass splitting between the neutral and the charged winos as a function of the neutral wino mass.
The figure shows that the two-loop contributions reduce the mass splitting by a few MeV 
compared to the central value of the one-loop result.
For $m_{\tilde\chi} = {\cal O}(1)~\TEV$, we find two-loop contribution is about $-2.8~\MEV$,
which can be understood as non-decoupling effect.
We can see that numerical value of mass splitting at two-loop level is consistent with the result of Ref. \cite{Yamada:2009ve}.
For $m_{\tilde \chi} \simeq 100~\GEV$, two-loop contribution is about $-3.5~\MEV$.
Then, we can see decoupling effect also diminishes wino mass splitting if wino mass is small,
although this effect is smaller than non-decoupling effect.
We also show the theoretical and experimental uncertainties as green/red bands.
As a result, we find that the uncertainties are significantly reduced by the two-loop analysis.
By numerical calculation, we have also  confirmed that the limit $m_{W,Z} \ll m_{\tilde\chi}$, 
our result reproduces the one in  Ref.\,\cite{Yamada:2009ve} at this level of precision
in the heavy wino limit, $M_2\gg m_{Z}$.

For the sake of readers, we give a fitting function of the central value of the two-loop result for $Q=\hat m_t$,
\begin{eqnarray}
\frac{\delta m}{1~\MEV} &=& -413.315
+305.383 \left(\log \frac{m_{\tilde\chi^0}}{1~\GEV}\right)
-60.8831 \left(\log\frac{m_{\tilde\chi^0}}{1~\GEV}\right)^2\nonumber\\
&& +5.41948 \left(\log\frac{m_{\tilde\chi^0}}{1~\GEV}\right)^3
-0.181509 \left(\log\frac{m_{\tilde\chi^0}}{1~\GEV}\right)^4.
\end{eqnarray}
for the central values of the SM input parameters.
Deviation of the above fitting function from our two-loop result is smaller than 0.02 \% for the wino mass being 100--4000 GeV.

%%%%%%%%%%%%%%%%%%%%%%%%%%%%%%%%%%%%%%%%%%%%%%%
%%%%%%%%%%%%%%%%%%%%%%%%%%%%%%%%%%%%%%%%%%%%%%%

%%%%%%%%%%%%%%%%%%%%%%%%%%%%%
\subsection{SUSY contributions}
\label{subsec: SUSY-particles}
Before closing this section, let us evaluate the contributions to the mass splitting from 
the diagrams including the heavy SUSY particles in the tens to hundreds TeV range.
Since the winos couple to the other gauginos (bino and gluinos) 
only through the exchange of those heavy particles, 
all the SUSY contributions to the wino masses can be expressed 
by the higher-dimensional operators suppressed by the heavy masses.
At a first glance, a five-dimensional operator, 
\begin{eqnarray}
{\cal L}_5 = \frac{1}{\Lambda}\epsilon_{abc} \tilde{\chi}^a \tilde{\chi}^b (H^\dagger \tau^c H)\ , 
\end{eqnarray}
with $H$ being the light Higgs boson and $\Lambda = {\cal O}(10$--$100)$\,TeV 
the scale of the heavy SUSY particles 
seems to break the custodial symmetry and contribute to the mass splitting. 
Here, the superscripts $a,b$ and $c$ denote the indices of the adjoint representation of SU(2)$_L$.
This operator, however, vanishes because of the Majorana nature of the winos, 
$\tilde{\chi}^a \tilde{\chi}^b = \tilde{\chi}^b \tilde{\chi}^a$. 
Another dimension-five operator  
\begin{eqnarray}
{\cal L}_5 = 
\frac{1}{\Lambda}\tilde{\chi}^a (H^\dagger \tau^a H)\tilde{b}\ , 
\label{eq:dim5-2}
\end{eqnarray}
with $\tilde b$ being the bino, on the other hand, contributes to the mass splitting
of ${\cal O}(v^4/\Lambda^2M_1)$ through the neutralino mass matrix.
Incidentally, the tree-level mass splitting due to the Higgsino
mixing in Eq.\,(\ref{eq: mixing}) can be regarded as one of the contributions 
of this type with $\Lambda \sim \mu$.
As a result, we find that the contributions from the dimension-five operators 
are negligibly small as we have seen in  Eq.\,(\ref{eq: mixing}).

The next lowest-dimensional operator which contributes  
to the mass splitting is the dimension-seven operator 
\begin{eqnarray}
{\cal L}_{\rm 7} =
\frac{M}{\Lambda^4}(\tilde{\chi}^a \tilde{\chi}^b)
(H^\dagger \tau^a H) (H^\dagger \tau^b H)\ ,
\label{eq:dim-7}
\end{eqnarray}
where $M$ denote the insertion of the gaugino mass.%
\footnote{
This operator can be obtained from, for instance,
a dimension-eight operator $(q_L \chi^a H)^\dagger(q_L \chi^a H)/\Lambda^4$ 
which is generated by integrating out the squarks (especially stops)
at the tree-level.
By integrating the quark-loop and inserting the gaugino mass,
we obtain the dimension-seven operator in Eq.\,(\ref{eq:dim-7}).
}
For $\Lambda = {\cal O}(10$--$100)$\,TeV, 
the contribution from this operator to the mass splitting is again negligibly small. 
%Therefore, in the models with the sfermions,  the Higgsino, the heavier Higgs bosons
%in the ${\cal O}(10$--$100$)\,TeV range, the mass splitting of the winos is dominated 
%by the contributions discussed in subsection\,\ref{subsec: SM-particles}.

%%%%%%%%%%%%%%%%%%%%
%%%%% Lifetime %%%%%
%%%%%%%%%%%%%%%%%%%%
\section{The charged wino decay} \label{sec: lifetime}
As we have seen in the previous section, the charged and the neutral winos are highly
degenerated in mass.
Therefore, the decay width of the charged wino is highly suppressed by the phase space integral,
and hence, the charged wino is long-lived and has the decay length about $c\tau = {\cal O}(1$--$10)$\,cm. 
With such a rather long decay length, it is possible to detect the charged wino production
at the LHC experiment by looking for disappearing tracks.
In this section, we estimate the lifetime of the charged wino and compare with the constraint 
from the disappearing track search by the ATLAS collaboration\,\cite{ATLAS:2012jp}.

With the small mass splitting $\delta m \sim 160\,\MEV$, 
the charged wino dominantly decays into a neutral wino and a soft charged pion.
%Its decay width can be calculated by using the partially conserved axial current (PCAC) relation. 
At the leading order, the decay width of the charged wino can be expressed in terms of 
the decay width of the charged pion,
\begin{eqnarray}
\G(\tilde \chi^\pm \to \tilde \chi^0 \pi^\pm)
=  \G( \pi^\pm \to \m^\pm \n_\m) \times \frac{16 \delta m^3}{m_\pi m_\m^2} \left( 1-\frac{m_\pi^2}{\delta m^2} \right)^{1/2} \left( 1-\frac{m_\m^2}{m_\pi^2} \right)^{-2}\ , \label{eq:topi}
\end{eqnarray}
where $m_{\pi}$ and $m_{\mu}$ denote the masses of the charged pion 
and the muon, respectively.\footnote{
At the next-to-leading order, Eq. (\ref{eq:topi}) receives radiative corrections
from the QED and the electroweak interactions which are expected to be around 
$(\a/\pi)\log(m_{\tilde\chi}/m_\pi) \simeq 2\,\%$.
In this Letter, we neglect these corrections to the total decay width and leave
the detailed analysis of the decay width for future study\,\cite{lifetime}.
}
%The branching ratio of the sub-leading leptonic decay mode into a pair of the electron and the
%neutrino\,\cite{Chen:1996ap},
The decay width of the sub-leading leptonic decay mode into a pair of the electron and the
neutrino\,\cite{Chen:1996ap} is given by
\begin{eqnarray}
\G(\tilde \chi^\pm \to \tilde \chi^0 e^\pm\n_e) &\simeq& \frac{2G_F^2}{15\pi^3} \delta m^5. \label{eq:toenu}
\end{eqnarray}
We consider the above two decay modes.

%%%%%%%%%%%%%%%%%%%%%%%%%%%%%%%%%%%%%%%%
\begin{figure}[p]
\begin{center}
\includegraphics[width=\hsize]{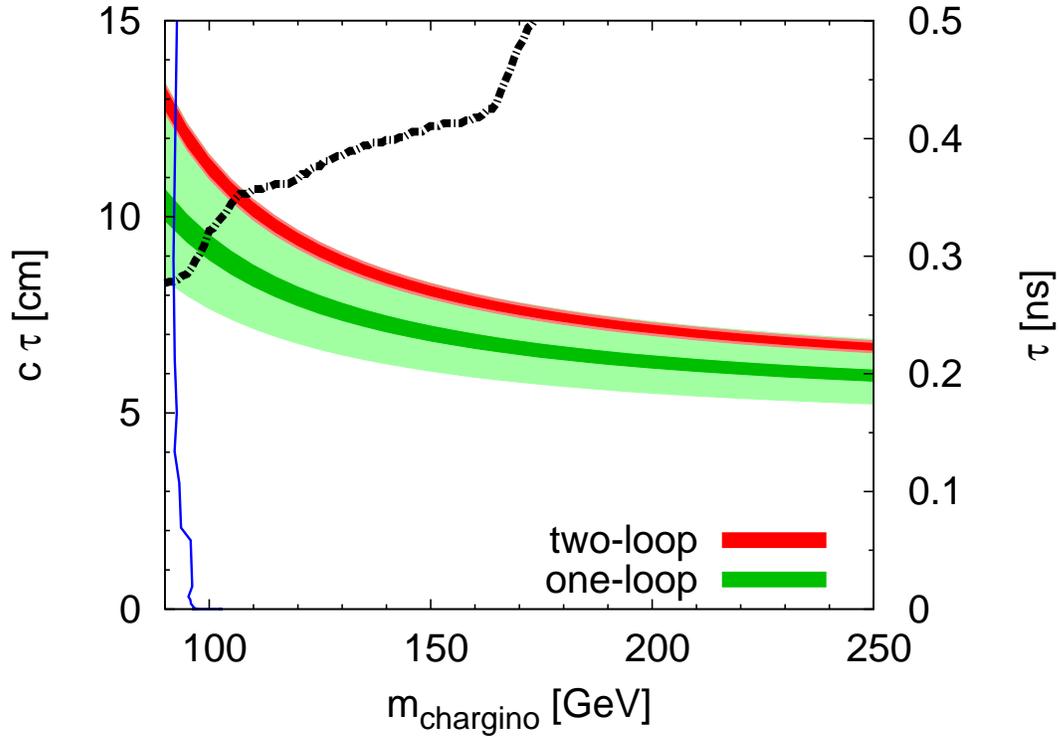}
\caption{
The lifetime of charged wino evaluated by using
$\delta m$ at the one-loop (green band) and two-loop (red band).
We neglected the next-to-leading order corrections 
to the lifetime of the charged wino estimated in terms of the pion decay rate,
which is expected to be a few percent correction.
%We also neglected the leptonic decay mode of the charged wino whose branching ratio is about 1\,\%.
The black chain line is the upper limit on the lifetime for a given
chargino mass  by the ATLAS collaboration at 95\,\%\,CL
($\sqrt{s}=7~\TEV$, ${\cal L}=4.7~{\rm fb}^{-1})$\,\cite{ATLAS:2012jp}.
The blue line shows the constraints which are given by the LEP2 constraints\,\cite{Heister:2002mn}--\cite{LEP2}.
}\label{fig:atlas}
\end{center}
\end{figure}
%%%%%%%%%%%%%%%%%%%%%%%%%%%%%%%%%%%%%%%%
In Fig.\,\ref{fig:atlas}, we show the lifetime of charged wino as a function of  the charged 
wino mass, $m_{\tilde\chi^\pm}$.
The meaning of the green and red bands are the same with the ones in Fig.\,\ref{fig:dm}.
The region above the black chain line is excluded by the disappearing charged track search 
by the ATLAS collaboration at $95$\,\% CL\,\cite{ATLAS:2012jp}.
The figure shows that the lifetime is enhanced by about 30\,\% for the wino mass around 100\,GeV
due to the two-loop contributions.
Furthermore, the figure also shows that 
the precise estimation of the mass splitting at the two-loop level
improves the constraint on the charged wino mass by about 10\,\%,
and the constraint by the ATLAS collaboration clearly exceeds the constraint by the LEP2 experiments\,\cite{Heister:2002mn}--\cite{LEP2},
which is shown as a blue line in Fig.\,\ref{fig:atlas}.

%%%%%%%%%%%%%%%%%%%
%%%%% Summary %%%%%
%%%%%%%%%%%%%%%%%%%
\section{Summary}
\label{sec: summary}
We have calculated the mass splitting of the charged and the neutral wino 
in the almost pure wino LSP scenario at the two-loop level.
Such a scenario attracts more attention after the discovery 
of the Higgs-like boson at the LHC experiment.
As a result, we found that the lifetime of the charged wino becomes
about 10--30\,\% longer due to the two-loop contributions.
Furthermore, we also found that the precise determination 
of the mass splitting improves the constraint on the mass of 
the wino obtained by the disappearing track search at the ATLAS
experiment by about 10\%.

%%%%%%%%%%%%%%%%%%%%%%%%%%%
%%%%% Acknowledgments %%%%%
%%%%%%%%%%%%%%%%%%%%%%%%%%%
\section*{Acknowledgments}
The authors thank T.\,T.\,Yanagida for useful discussions at the early stage of the collaboration.
This work is supported by the Grant-in-Aid for Scientific research from the Ministry of Education, Science, Sports, and Culture (MEXT), Japan (No. 24740151 for M.I., Nos. 22244021, 23740169 for S.M.), and also by the World Premier International Research Center Initiative (WPI Initiative), MEXT, Japan. The work of R.S. is supported in part by JSPS Research Fellowships for Young Scientists.

%%%%%%%%%%%%%%%%%%%%%%%%%%%%%%%%%%
%%%%%%%%%%% Appendices %%%%%%%%%%%
%%%%%%%%%%%%%%%%%%%%%%%%%%%%%%%%%%
\appendix

%%%%%%%%%%%%%%%%%%%%%%%%%%
%%%%% Loop functions %%%%%
%%%%%%%%%%%%%%%%%%%%%%%%%%
\section{Loop functions}
\label{sec: loop-functions}
The one-loop functions are presented in this appendix, which are used to calculate radiative corrections to the self-energies of SM particles and winos:
{\small
\begin{eqnarray}
B_0(p^2, m_1^2, m_2^2) &=& \Delta
- \int_0^1 dx ~\log \frac{ (1-x)m_1^2 + x m_2^2 - x(1-x)p^2 -i\epsilon }{Q^2}\ , \\
B_1(p^2, m_1^2, m_2^2) &=& -\frac{\Delta}{2}
+ \int_0^1 dx ~x\log \frac{ (1-x)m_1^2 + x m_2^2 - x(1-x)p^2 -i\epsilon }{Q^2}\ , \\
B_{21}(p^2, m_1^2, m_2^2) &=& \frac{\Delta}{3}
- \int_0^1 dx ~x^2\log \frac{ (1-x)m_1^2 + x m_2^2 - x(1-x)p^2 -i\epsilon }{Q^2}\ ,
\end{eqnarray}
}where $\Delta = 2/(4-d) - \gamma_E + \log(4\pi)$.  With $M$ being much larger than $m$, the functions $B_0(M^2, M^2, m^2)$ and $B_1(M^2, M^2, m^2)$ are evaluated as follows;
{\small
\begin{eqnarray}
B_0(M^2, M^2, m^2) &=& \Delta
- \log\frac{M^2}{Q^2} + 2 - \frac{\pi m}{M}
+ {\cal O}\left( \frac{m^2}{M^2} \log\frac{M^2}{m^2} \right)\ , \\
B_1(M^2, M^2, m^2) &=& -\frac{\Delta}{2}
+ \frac{1}{2}\log\frac{M^2}{Q^2} - \frac{3}{2} + \frac{\pi m}{M}
+ {\cal O}\left( \frac{m^2}{M^2} \log\frac{M^2}{m^2} \right)\ .
\end{eqnarray}
}
By using $B_0(p^2, m_1^2, m_2^2)$, $B_1(p^2, m_1^2, m_2^2)$, and $B_{21}(p^2, m_1^2, m_2^2)$, 
we define $\Pi_V(p^2, m_1^2, m_2^2)$ and $\tilde B_{22}(p^2, m_1^2, m_2^2)$ by
{\small
\begin{eqnarray}
\Pi_V(p^2, m_1^2, m_2^2) &=&
-p^2 [B_1(p^2,m_1^2,m_2^2) + B_{21}(p^2,m_1^2,m_2^2) ]\ , \\
{\tilde B}_{22}(p^2, m_1^2, m_2^2) &=&
- p^2 (B_1 + B_{21}) - \frac{p^2}{4} B_0 - \frac{1}{4}(m_1^2 - m_2^2) (B_0 + 2B_1)\ .
\end{eqnarray}
}
%%%%%%%%%%%%%%%%%%%%%%%%%%%%%%
%%%%% 1-loop corrections %%%%%
%%%%%%%%%%%%%%%%%%%%%%%%%%%%%%
\section{Radiative corrections at one-loop}
\label{sec: one-loop-corrections}
Here, all radiative corrections to the 1PI self-energies of the gauge bosons and the winos 
at one-loop level are presented. 
The counter-terms (in $\overline{\rm MS}$ scheme) to eliminate the one-loop UV divergences  
are also shown. 
These self-energies as well as the counter-terms are used in the calculation of the mass splitting at two-loop level.
We have checked that self-energies which are given in this appendix are consistent with Ref. \cite{Yamada:2009ve} and Ref. \cite{Pierce:1996zz}.\footnote{
In our notation, sign of $\Pi_{\g Z}$ is opposite to Refs. \cite{Yamada:2009ve, Pierce:1996zz}.
We have calculated self-energies in ${\overline{\rm MS}}$ scheme, then,
our calculation does not include the contribution of $\e$-scalar unlike Ref. \cite{Pierce:1996zz}.
}

\subsection{Gauge boson self-energies}
\label{sec: oblique}
In terms of the 1PI amplitude $\Pi(p^2)$, the full propagator (2-point function) of the gauge boson in the Feynman gauge is given by
$(-ig_{\mu\nu})/[p^2 - \hat{m}_V^2 + \Pi(p^2)]$. 
In this subsection, we present the contributions to the amplitude 
$\Pi(p^2)$ from both the SM particles and the winos, 
which are divided into three parts;
{\small
\begin{eqnarray}
\Pi_{V_1 V_2}
= \Pi_{V_1 V_2}^{(q, \ell)}
+ \Pi_{V_1 V_2}^{(V, h)}
+ \Pi_{V_1 V_2}^{(\tilde{\chi})}
+ p^2 \delta_{Z_{V_1 V_2}}
+ \delta_{m^2_{V_1 V_2}}\ ,
\end{eqnarray}
}where $V_1 V_2 =$ $\gamma \gamma$, $\gamma Z$, $ZZ$, and $WW$. The first term in the right-hand side is the contributions from the quarks and the leptons, the second term is those from the gauge-Higgs sector of the SM, and the third term is from the neutral and charged winos. 
The fourth and fifth terms show the counter-terms given in appendix \ref{sec: counter}.

\subsubsection{Contributions from winos}
{\small
\begin{eqnarray}
\Pi_{\gamma \gamma}^{(\tilde{\chi})}(p^2) &=&
\frac{\hat{e}^2}{2\pi^2}
\Pi_V(p^2, \hat{M}_2^2, \hat{M}_2^2)\ , \\
%%%%%
\Pi_{\gamma Z}^{(\tilde{\chi})}(p^2) &=&
-\frac{\hat{e} \hat{g} \hat{c}_W}{2\pi^2}
\Pi_V(p^2, \hat{M}_2^2, \hat{M}_2^2)\ , \label{eq:winoloop1}\\
%%%%%
\Pi_{ZZ}^{(\tilde{\chi})}(p^2) &=&
\frac{\hat{g}^2 \hat{c}_W^2}{2\pi^2}
\Pi_V(p^2, \hat{M}_2^2, \hat{M}_2^2)\ , \\
%%%%%
\Pi_{WW}^{(\tilde{\chi})}(p^2) &=&
\frac{\hat{g}^2}{2\pi^2}
\Pi_V(p^2, \hat{M}_2^2, \hat{M}_2^2)\ . \label{eq:winoloop2}
\end{eqnarray}
}
With the use of the above amplitudes, the finite renormalization effect, $\tilde{\Pi}_{\gamma \gamma}^{(\tilde{\chi})}(p^2)$, in Eq.\,(\ref{eq: alpha-MS-bar}) is given by the combination, $\tilde\Pi_{\gamma \gamma}^{(\tilde{\chi})}(p^2) = \Pi_{\gamma \gamma }^{(\tilde{\chi})}(p^2) - p^2 (\hat{e}^2/12\pi^2) \Delta$.

\subsubsection{Contributions from quarks and leptons}

{\small
\begin{eqnarray}
\Pi_{\gamma \gamma}^{(q, \ell)}(p^2) &=&
\sum_f \frac{\hat{e}^2 N^C_f}{2\pi^2} Q_f^2
\Pi_V(p^2, \hat{m}_f^2, \hat{m}_f^2)\ , \\
%%%%%
\Pi_{\gamma Z}^{(q, \ell)}(p^2) &=&
-\sum_f \frac{\hat{e} \hat{g} N^C_f}{2\pi^2 \hat{c}_W} Q_f Z_f
\Pi_V(p^2, \hat{m}_f^2, \hat{m}_f^2)\ , \label{eq:fermionloop1}\\
%%%%%
\Pi_{Z Z}^{(q, \ell)}(p^2) &=&
\sum_f \frac{\hat{g}^2 N^C_f}{2\pi^2 \hat c_W^2}
\Biggl[\left(\frac{T_f^2}{4} + Z_f^2\right) \Pi_V(p^2, \hat{m}_f^2, \hat{m}_f^2) %\nonumber\\
%&&
%~~~~~~~~~~~~~~~~~~~~~~~~~~~~~~~~~~~
+  \frac{T_f^2}{4} \hat{m}_f^2 B_0(p^2, \hat{m}_f^2, \hat{m}_f^2) \Biggr], \\
%%%%%
\Pi_{WW}^{(q, \ell)}(p^2) &=&
\sum_{f_u/f_d} \frac{\hat{g}^2 N^C_f}{8\pi^2}
\Biggl[
\Pi_V(p^2, \hat{m}_u^2, 0) 
%\Pi_V(p^2, \hat{m}_u^2, \hat{m}_d^2) \nonumber\\
%&&
%~~~~~~~~~~~~~
+\frac{\hat{m}_u^2}{2}\left[
 B_0(p^2, \hat{m}_u^2, 0) + B_1(p^2, \hat{m}_u^2, 0)
%\hat{m}_u^2 B_0(p^2, \hat{m}_u^2, \hat{m}_d^2) + (\hat{m}_u^2-\hat{m}_d^2) B_1(p^2, \hat{m}_u^2, \hat{m}_d^2)
\right] \Biggr]\ ,\label{eq:fermionloop2}
\end{eqnarray}
}where $Q_f$ is the electric charge of the fermion $f$, while $T_f$ takes the value $1/2$ and $-1/2$ for up-type fermions ($u$, $c$, $t$ quarks and neutrinos) and down-type fermions ($d$, $s$, $b$ quarks and charged leptons), respectively. The coefficient $Z_f$ is given by the equation $Z_f = (T_f/2 - Q_f \hat{s}_W^2)$, while $N^C_f =3$  for the quarks and $N^C_f =1$ for the leptons. The summation $\sum_{f_u/f_d}$ should be over left-handed quarks and leptons. The finite mass effect of the down-type fermions on the amplitude $\Pi_{WW}^{(q, \ell)}(p^2)$ is neglected.

\subsubsection{Contributions from the gauge-Higgs sector}

{\small
\begin{eqnarray}
\Pi_{\gamma \gamma}^{(V, h)}(p^2) &=&
-\frac{3 \hat{e}^2}{4\pi^2}
\left[ \tilde{B}_{22}(p^2, \hat{m}_W^2, \hat{m}_W^2) + \frac{p^2}{18} \right]
-\frac{\hat{e}^2 p^2}{4\pi^2} B_0(p^2, \hat{m}_W^2, \hat{m}_W^2)\ , \\
%%%%%
\Pi_{\gamma Z}^{(V, h)}(p^2) &=&
\frac{\hat{e} \hat{g}}{8\pi^2 \hat{c}_W}( 6\hat{c}_W^2 - 1)
\tilde{B}_{22}(p^2, \hat{m}_W^2, \hat{m}_W^2)
+\frac{\hat{e} \hat{g} \hat{c}_W p^2}{24\pi^2} \nonumber \\
&& +\frac{\hat{e} \hat{g}}{8\pi^2 \hat{c}_W}(2\hat{c}_W^2 p^2 + \hat{m}_W^2)
B_0(p^2, \hat{m}_W^2, \hat{m}_W^2)\ , \label{eq:vhloop1}\\
%%%%%
\Pi_{ZZ}^{(V,h)}(p^2) &=& 
-\frac{ \hat{g}^2 (12\hat{c}_W^4 - 4\hat{c}_W^2 + 1) }{16\pi^2 \hat{c}_W^2}
\tilde B_{22}(p^2, \hat{m}_W^2, \hat{m}_W^2) 
-\frac{ \hat{g}^2 \hat{c}_W^2 p^2 }{24\pi^2} \nonumber \\
&& -\frac{2\hat{g}^2}{16\pi^2}(2\hat{c}_W^2 p^2 + 2\hat{m}_W^2 - \hat{m}_Z^2)
B_0(p^2, \hat{m}_W^2, \hat{m}_W^2) \nonumber \\
&& -\frac{ \hat{g}^2 }{16\pi^2 \hat{c}_W^2}
[\tilde B_{22}(p^2, \hat{m}_Z^2, \hat{m}_h^2)
-\hat{m}_Z^2 B_0(p^2, \hat{m}_Z^2, \hat{m}_h^2) ]\ , \\
%%%%%
\Pi_{WW}^{(V, h)}(p^2) &=&
-\frac{8\hat{e}^2}{16\pi^2} \tilde B_{22}(p^2, 0, \hat{m}_W^2)
-\frac{\hat{e}^2 p^2}{24\pi^2}
-\frac{4\hat{e}^2 p^2}{16\pi^2} B_0(p^2, 0, \hat{m}_W^2) \nonumber \\
&& -\frac{\hat{g}^2}{16\pi^2}(1 + 8\hat{c}_W^2)
\tilde{B}_{22}(p^2, \hat{m}_W^2, \hat{m}_Z^2)
-\frac{\hat{g}^2 \hat{c}_W^2 p^2}{24\pi^2} \nonumber \\
&& -\frac{\hat g^2}{16\pi^2}( 4 \hat c_W^2 p^2 + 3 \hat m_W^2 - \hat m_Z^2) B_0( p^2, \hat m_W^2, \hat m_Z^2) \nonumber\\
&& -\frac{\hat{g}^2}{16\pi^2}
[ \tilde{B}_{22}(p^2, \hat{m}_W^2, \hat{m}_h^2)
-\hat{m}_W^2 B_0(p^2, \hat{m}_W^2, \hat{m}_h^2) ]\ .\label{eq:vhloop2}
\end{eqnarray}
}

\subsection{Wino self-energies}
\label{sec: wino-one-loop}
With the use of the 1PI amplitudes $\Sigma_K(p^2)$ and $\Sigma_M(p^2)$, the full propagators (2-point functions) of the winos are given by $i/[\{1+\Sigma_K(p^2)\}\slashed{p} - \hat{M}_2 + \Sigma_M(p^2)]$. In this subsection, we explicitly present the amplitudes for both neutral and charged winos at the 
one-loop level. For the neutral wino, the amplitudes are given by
{\small
\begin{eqnarray}
\Sigma_{K, 0}^{(1)} &=&
-\frac{\hat{g}^2}{16\pi^2}
\left[ 4 B_1(p^2, \hat{M}_2^2, \hat{m}_W^2) + 2 \right]
+\delta_{Z_{\tilde{\chi}}}\ , \label{eq: Neu_K} \\
%%%%%
\Sigma_{M, 0}^{(1)} &=&
-\frac{\hat{g}^2 \hat{M}_2}{16\pi^2}
\left[ 8 B_0(p^2, \hat{M}_2^2, \hat{m}_W^2) - 4 \right]
-\delta_{M_{\tilde{\chi}}}\ . \label{eq: Neu_M}
\end{eqnarray}
}On the other hand, the two amplitudes for the charged wino are given by
{\footnotesize
\begin{eqnarray}
\Sigma_{K, \pm}^{(1)} =
-\frac{\hat{g}^2}{8\pi^2}
\left[ \hat{s}_W^2 B_1(p^2,\hat{M}_2^2,0)
+\hat{c}_W^2 B_1(p^2,\hat{M}_2^2, \hat m_Z^2)
+B_1(p^2, \hat{M}_2^2, \hat{m}_W^2) + 1  \right]
+ \delta_{Z_{\tilde{\chi}}}\ , \label{eq: Cha_K} \\
%%%%%
\Sigma_{M, \pm}^{(1)} =
-\frac{\hat{g}^2 \hat{M}_2}{4\pi^2}
\left[ \hat{s}_W^2 B_0(p^2, \hat{M}_2^2,0)
+\hat{c}_W^2 B_0(p^2, \hat{M}_2^2, \hat{m}_Z^2)
+B_0(p^2, \hat{M}_2^2, \hat{m}_W^2) - 1 \right]
-\delta_{M_{\tilde{\chi}}}\ ,
 \label{eq: Cha_M}
\end{eqnarray}
}where explicit forms of the counter-terms, $\delta_{Z_{\tilde{\chi}}}$ and $\delta_{M_{\tilde{\chi}}}$, are given in Appendix \ref{sec: counter}.

\subsection{Counter-terms}
\label{sec: counter}
Finally, we  give the counter-terms in the $\overline{\rm MS}$ scheme in the framework of the SM plus the winos. These are used in the calculations of the self-energies mentioned above and of the mass splitting at the two-loop level as shown in Fig.\,\ref{fig: 1-loop-ct}.

\subsubsection{Gauge boson self-energies}

{\small
\begin{eqnarray}
\delta_{Z_{\gamma \gamma}} &=&
-\frac{\hat{e}^2}{16\pi^2}
\left(\frac{32}{9} N_g - \frac{5}{3} \right)\Delta\ , \\
%%%%%
\delta_{Z_{\gamma Z}} &=&
-\frac{\hat{e} \hat{g}}{16\pi^2 \hat{c}_W}
\left[
\left(-\frac{4}{3} + \frac{32 \hat{s}_W^2}{9} \right) N_g
+ \left(\frac{11}{6} - \frac{5\hat{s}_W^2}{3} \right)
\right]\Delta\ , \\
%%%%%
\delta_{Z_{Z Z}} &=&
-\frac{\hat{g}^2}{16\pi^2 \hat{c}_W^2}
\left[
\left(\frac{4}{3} - \frac{8}{3}\hat{s}_W^2 + \frac{32}{9}\hat{s}_W^4 \right) N_g
+ \left(-\frac{11}{6} + \frac{11}{3}\hat{s}_W^2 - \frac{5}{3} \hat{s}_W^4 \right)
\right]\Delta\ , \\
%%%%%
\delta_{Z_{W W}} &=&
-\frac{\hat{g}^2}{16\pi^2}
\left(\frac{4}{3} N_g - \frac{11}{6} \right)\Delta\ ,
\end{eqnarray}
} where $N_g$ is the number of the generation, namely $N_g = 3$ for the SM.
{\small
\begin{eqnarray}
\delta_{m^2_{\gamma Z}} &=&
-\frac{\hat{e} \hat{g}}{16\pi^2 \hat{c}_W} (2-2\hat{s}_W^2)\hat{m}_Z^2
\Delta\ , \\
%%%%%
\delta_{m^2_{Z Z}} &=&
-\frac{\hat{g}^2}{16\pi^2 \hat{c}_W^2}
\left[-\frac{3}{2}\hat{m}_t^2
+(-1 + 6\hat{s}_W^2 - 4\hat{s}_W^4) \hat{m}_Z^2 \right]
\Delta\ , \\
%%%%%
\delta_{m^2_{WW}} &=&
-\frac{\hat{g}^2}{16\pi^2}
\left[-\frac{3}{2}\hat{m}_t^2 + (-1 + 2\hat{s}_W^2) \hat{m}_Z^2 \right]
\Delta\ ,
\end{eqnarray}
}where we have neglected the masses of all the SM fermions except the top quark.

\subsubsection{Wino self-energies}

\begin{eqnarray}
\delta_{Z_{\tilde{\chi}}} &=& -\frac{\hat{g}^2}{8\pi^2} \Delta\ , \\
\delta_{M_{\tilde{\chi}}} &=& -\frac{\hat{g}^2 \hat{M}_2}{2\pi^2} \Delta\ .
\end{eqnarray}

\subsubsection{Gauge interaction of the wino}

The neutral and charged winos have the SU(2)$_L$ gauge interaction which is described by the term, ${\cal L}_{\rm int} = i\epsilon_{abc}(\hat{g} + \delta_{\tilde{\chi} \tilde{\chi} W}) \tilde{\chi}^{a \dagger} \slashed{W}^b \tilde{\chi}^c$, and the counter-term is given by
\begin{eqnarray}
\delta_{\tilde{\chi} \tilde{\chi} W} = \frac{\hat{g}^3}{4\pi^2} \Delta\ .
\end{eqnarray}

\section{Cancellation of IR divergences}
\label{app: IR cancel}
In Eq.\,(\ref{eq: pole_mass}) with $M_0$ being $\hat{M}_2$, 
the IR divergences appear in $[(\Sigma_M^{(1)} + M_0 \Sigma_K^{(1)}) (2M_0\dot{\Sigma}_M^{(1)} + 2M_0^2 \dot{\Sigma}_K^{(1)} )]_{p^2 = M_0^2}$ and $(-1) (\Sigma_M^{(2)} + M_0 \Sigma_K^{(2)})$. 
The first term is the products of the one-loop contributions.
The one-loop amplitude $(\Sigma_M^{(1)} + M_0 \Sigma_K^{(1)})$ is explicitly written as
{\small
\begin{eqnarray}
F_{\rm 1L}(p^2) \equiv \Sigma_M^{(1)}(p^2) + M_0 \Sigma_K^{(1)}(p^2) =
\int \frac{d^4 k}{(2\pi)^4}
\frac{ (ie^2) \gamma^\mu ( \slashed{k} + \slashed{p} + M_0) \gamma_\mu }
{[k^2 - m_\gamma^2] [(k + p)^2 - M_0^2]}
+ \cdots\ ,
\label{eq: one-loop-IR}
\end{eqnarray}
}where we have introduced a photon mass $m_\gamma$ to control the IR divergences.
The ellipses stand for the contributions from the loop diagrams of the $W$ and $Z$ bosons, 
which are nothing to do with the IR divergences. 
The derivative of the one-loop amplitude $F_{\rm 1L}(q^2)$  with respect to $p^2$ 
gives the IR-divergent contribution,
{\small
\begin{eqnarray}
\frac{d}{dp^2} F_{\rm 1L}(p^2) =
\int \frac{d^4 k}{(2\pi)^4}
\frac{(-ie^2) 2M_0 \gamma^\mu \gamma_\mu }
{[k^2 - m_\gamma^2][(k + p)^2 - M_0^2]^2}
+ \cdots\ ,
\label{eq: one-loop-diff-infra}
\end{eqnarray}
}where the ellipses represent the terms which do not cause the IR divergences, 
namely the IR-safe terms. 

The second contribution, $F_{\rm 2L} \equiv (-1) (\Sigma_M^{(2)} + M_0 \Sigma_K^{(2)})$, 
is, on the other hand, written as
{\small
\begin{eqnarray}
F_{\rm 2L}(M_0^2) =
(i e^2) \int \frac{d^4 k}{(2\pi)^4}
\frac{
\gamma^\mu (\slashed{k} + M_0)
[\Sigma_K^{(1)}(k^2) \slashed{k} + \Sigma_M^{(1)}(k^2)]
(\slashed{k} + M_0)\gamma_\mu
}
{[ (k-p)^2 - m_\gamma^2 ] [ k^2 - M_0^2 ]^2 } \Biggr|_{p^2 = M_0^2}
+ \cdots\ .
\end{eqnarray}
}The numerator of the integrand in above equation can be simplified as
{\small
\begin{eqnarray}
&&
\gamma^\mu
2M_0 [M_0 \Sigma_K^{(1)}(k^2) + \Sigma_M^{(1)}(k^2)] (\slashed{k} + M_0)
\gamma_\mu
+ {\cal O}(k^2 - M_0^2)
\nonumber \\
&=&
2M_0 [M_0 \Sigma_K^{(1)}(M_0^2) + \Sigma_M^{(1)}(M_0^2)] 2M_0
\gamma^\mu \gamma_\mu
+ {\cal O}(k^2 - M_0^2)\ . 
\end{eqnarray}
}
As a result, the IR-divergent part of the two-loop contribution can be reduced to,
{\small
\begin{eqnarray}
F_{\rm 2L}(M_0^2) =
(2 M_0) \left[M_0 \Sigma_K^{(1)}(M_0^2) + \Sigma_M^{(1)}(M_0^2)\right]
\times
\left. \frac{d}{dp^2} F_{\rm 1L}(p^2) \right|_{p^2=M_0^2}
+ \cdots\ .
\label{eq: two-loop-infra}
\end{eqnarray}
}
Therefore,
we find that the IR-divergences cancel with each other,
and hence, the pole mass is an IR-safe quantity.

\newpage

%%%%%%%%%%%%%%%%%%%%%%
%%%%% References %%%%%
%%%%%%%%%%%%%%%%%%%%%%
%\input{reference}

\end{document}